\documentclass[vecphys]{svmult}

\usepackage{makeidx}         
\usepackage{graphicx}        
\usepackage{multicol}        
\usepackage[bottom]{footmisc}

\makeindex             

\begin{document}

\title*{Globular clusters with the extended horizontal-branch as remaining cores of galaxy building blocks}
\author{Young-Wook Lee, Hansung B. Gim, \and
Chul Chung}
\authorrunning{Lee, Gim, \& Chung}
\titlerunning{EHB GCs as remaining cores of galaxy building blocks}
\institute{{Center for Space Astrophysics, Yonsei University, Seoul 120-749, Korea}
\texttt{ywlee@csa.yonsei.ac.kr}}

\maketitle

\begin{abstract}

  The relics of building blocks that made stellar halo and bulge are yet to be discovered unless 
they were completely disrupted throughout the history of the Galaxy. Here we suggest that about 25\% 
of the Milky Way globular clusters have characteristics of the remaining cores of these early building blocks 
rather than genuine star clusters. They are clearly distinct from other normal globular clusters 
in the presence of extended horizontal-branch and multiple stellar populations, in mass (brightness), 
and most importantly in orbital kinematics. Based on this result, a three-stage formation picture of the 
Milky Way is suggested, which includes early mergers, collapse, and later accretion.

\end{abstract}

\section{Introduction}
\label{sec:1}

  Our view of the Milky Way Galaxy formation has been changing since the early pioneering work by 
Eggen, Lynden-Bell, \& Sandage (1962, hereafter ELS), where it was suggested that the Galaxy formed 
by the monolithic collapse of a gas-rich proto-Galaxy. Searle \& Zinn (1978), Toomre (1977), 
and other investigators then suggested that the Galaxy formed more likely by accretion of many gas-rich subsystems. 
By 1993, these two different pictures were merged together, and it was suggested that while the 
outer halo formed mostly by accretion, the inner halo formed mainly by collapse much like 
that originally suggested by ELS ([4], [5]). Since then, more evidence was 
discovered for the accretion, such as Sagittarius dwarf galaxy [6] and halo substructures, 
along with the recent progress in theoretical simulations. In particular, according to the currently favored 
$\Lambda$CDM hierarchical merging paradigm, a galaxy like the Milky Way formed by numerous mergers of smaller subsystems. 
However, most of these building blocks are yet to be discovered or identified. Where are 
the relics of building blocks that formed stellar component of the Galaxy ?

\section{Clues from $\omega$ Cen and other massive globular clusters}
\label{sec:2}
  Freeman (1993), among others, suggested that some globular clusters (GCs) might be remaining cores of 
disrupted nucleated dwarf galaxies. This idea was strengthened by the discovery of multiple stellar 
populations from the giant-branch of $\omega$ Cen [8], which together with 
other works have basically established that $\omega$ Cen is not a genuine GC, 
but a remaining core of a disrupted dwarf galaxy. More recent observations of $\omega$ Cen 
have also found equally curious double main-sequences (MS), with a minority population of bluer 
and fainter MS well separated from the majority of redder and brighter MS [9]. 
This has provided further evidence for the presence of multiple populations in this GC, 
although the observed feature was not easily reproduced from the standard assumptions on chemical compositions. 
Soon it was suggested by Norris (2004) that the presence of helium enhanced subpopulations, 
in addition to the majority of normal helium population, can best reproduce the observed features on the MS. 
Lee et al. (2005) then showed that the same helium enhancement required to reproduce 
this special feature on the MS was also able to reproduce the presence of extremely blue HB 
in the same cluster without further fine-tuning of the parameters, such as mass-loss on the giant-branch. 
So, in $\omega$ Cen, we have a case where one assumption (helium enhancement) can 
naturally explain two peculiar observations simultaneously, which, 
as we know from the history of science, then might be the case !

  Furthermore, the same scenario can also explain the strongly extended HB in other peculiar GCs, 
such as NGC 2808 and NGC6388/6441 ([11], [12]). 
These authors have specifically predicted that the MS of these GCs would also be splitted or broadened, 
and the HST/ACS is now confirming that the GCs with extended HB (EHB) are 
indeed showing either splitted or broadened MS [13]. 
This would ensure that EHB is a strong signature of the presence of multiple populations in a GC.

  Unfortunately, the origin of this helium enrichment is not fully understood yet. 
However, a significant fraction ($\sim 30\%)$ of the helium enriched subpopulation observed in these peculiar 
GCs appears to be best explained, if the second generation stars were formed from enriched gas trapped 
in the deep potential well while these GCs were cores of the ancient dwarf galaxies [14]. 
All of these recent developments suggest that GCs with extended HB (``EHB GCs'') are probably not genuine GCs, 
but might be remaining cores or relics of ancient building blocks of the Galaxy. 
Strong support for this possibility has recently been provided by Lee et al. (2007), 
who showed that EHB GCs are clearly distinct from GCs with normal HB in orbital kinematics and total stellar mass.\\

\section {Evidence from luminosity function and kinematics}
\label{sec:3}

\begin{figure} [!h]
  \centering
  \includegraphics[scale=0.87] {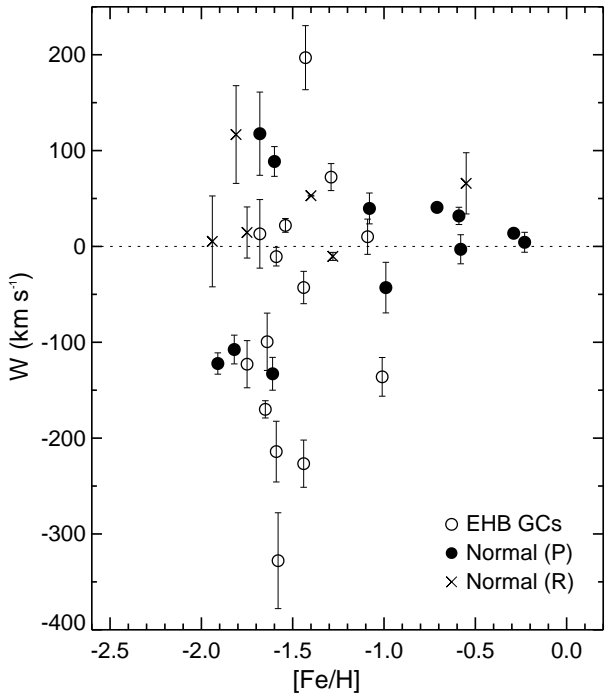}
  \caption{The correlations between kinematics (W velocity) and metallicity for GCs with and without
EHB in the ``old halo'' group, which are preferentially located in the inner halo.
Note the clear difference between GCs with and without EHB.}
  \label{fig:1}
\end{figure}

  About 25\% of the Milky Way GCs exhibit unusually extended HB [15]. 
Interestingly, these EHB GCs are clearly brighter (massive) than normal GCs, 
including 11 of the 12 brightest GCs (see Fig. 1 of [15]). 
Their inferred stellar mass, which is very likely to be only small fraction of their original mass, 
is already comparable to low-luminosity dwarf galaxies in the Local Group. 
Perhaps, this is already telling us that the origin of EHB GCs is indeed unique!

\begin{figure} [t]
  \centering
  \includegraphics[scale=0.51, angle=270] {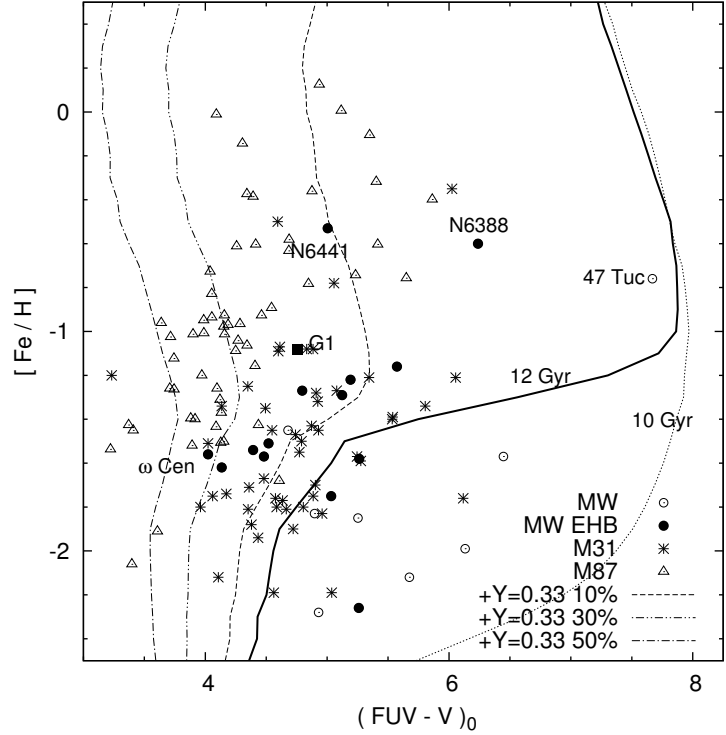}
  \begin{minipage}[t]{6.5cm}
    \vspace{5mm}
    \vspace{2mm}{\scriptsize \hspace{32mm}}
  \end{minipage}
  \caption{(FUV-V) integrated color plotted against metallicity for GCs in two disk galaxies (Milky Way and M31)
compared with those in gE galaxy M87 (data from [16] \& [17]). The solid and dotted lines 
are our model predictions for normal GCs without EHB (at 10 and 12 Gyrs). Other model lines are for EHB GCs (at 12 Gyr) 
with different fraction of helium enhanced subpopulation. Compared at given metallicity, GCs in gE galaxy appear to be 
bluer than those in disk galaxies.}
  \label{fig:2}
\end{figure}

  More interesting finding is, however, that the orbital kinematics of EHB GCs appears to be clearly 
decoupled from normal GCs ! For example, we have plotted in Figure 1 the correlation between metallicity 
and W velocity, a velocity component perpendicular to the Galactic plane, for GCs in ``old halo'' group, 
including metal-rich ``disk-bulge'' GCs. These classifications are, following Zinn (1993), 
based on the HB morphology versus metalliicty plane (see Fig. 2 of [15]), 
but note that most GCs in Figure 1 are those in the inner halo (R $<$ 8 kpc). 
We can see in Figure 1 that EHB GCs show diversity in kinematics but no correlation with metallicity, 
as would be expected among the relics of ancient subsystems that first assembled to form the nucleus and 
halo of the Galaxy. Surprisingly, however, when they are excluded, most normal GCs, 
especially those with prograde rotation (filled circles), show very special ``chevron'' shape distribution. 
If we randomly choose a group of GCs from total sample, we will hardly end up with this kind of unique distribution. 
In fact, Monte-Carlo simulations suggest that the possibility of getting this only by chance is less than $0.001\%$ ! 
This and the coherent behaviors of other kinematic parameters considered by Lee et al (2007; see their Fig. 3) 
appear to be fully consistent with the ELS type dissipational collapse that eventually led to the formation of the disk. 
Unlike GCs in the inner halo, normal GCs in the outer halo share their kinematic properties 
with the outlying EHB GCs (see Fig. 3 of [15]), which is consistent with the accretion origin of the outer halo.

\section {Discussion}
\label{sec:4}
  All of these differences between GCs with and without EHB are fully consistent with the idea 
that present-day Galactic GCs are ensemble of heterogeneous objects originated from three distinct phases 
of the Milky Way formation. Possible scenario is 
that (1) early mergers of building blocks (ancient dwarf galaxies, ``rare peaks'', 
gas-rich subsystems) forming the nucleus and halo of the proto-Galaxy, 
leaving their dense cores as today's EHB GCs, (2) formation of normal GCs 
in the dissipational collapse of a transient gas-rich inner halo system that eventually formed the Galactic disk, 
and finally (3) accretion of normal GCs originally formed in the outskirts of 
outlying building blocks to the outer halo of the Milky Way.

According to this picture, normal and genuine GCs formed in phase (2) would not be present 
in most giant elliptical galaxies (gEs), if the major star formation in these galaxies 
occurred before the formation of disk in subsystems (or building block galaxies) that merged to form gEs. 
If so, we would expect that the fraction of EHB GCs would be much higher among GCs in gEs, 
especially in the inner halo. Their mean integrated colors, especially in far-UV (FUV), 
would then be bluer, at given metallicity, than GCs in disk galaxies like the Milky Way. 
Interestingly, some support for this is provided from the far-UV observations of GCs in M87 (see Fig. 2). 
Further study will certainly help to shed more light into the new picture envisioned here.



\printindex

\begin{thebibliography}{99.}

\bibitem{journal} O.J. Eggen, D. Lynden-Bell, \& A.R. Sandage: ApJ \textbf{136}, 748 (1962)

\bibitem{journal} L. Searle \& R. Zinn: ApJ \textbf{225}, 357 (1978)

\bibitem{contribution} A. Toomre: Mergers and Some Consequences. In: \textit{Evolution of 
Galaxies and Stellar Populations}, ed by B.M. Tinsley \& R.B. Larson (Yale University Observatory, 
New Haven 1977) pp 401--426

\bibitem{contribution} R. Zinn: The Galactic Halo Cluster Systems: Evidence for Accretion. In: \textit{
The globular clusters-galaxy connection: ASP Conf. Ser.}, vol 48, ed by G.H. Smith and J.P. Brodie 
(ASP, San Francisco 1993) pp 38--47

\bibitem{journal} S. van den Bergh: ApJ \textbf{411}, 178 (1993)

\bibitem{journal} R.A. Ibata, G. Gilmore, \& M.J. Irwin: Nature \textbf{370}, 194 (1994)

\bibitem{contribution} K.C. Freeman: Globular Clusters and Nucleated Dwarf Ellipticals. In: \textit{
The globular clusters-galaxy connection: ASP Conf. Ser.}, vol 48, ed by G.H. Smith and J.P. Brodie 
(ASP, San Francisco 1993) pp 608--614

\bibitem{journal} Y.-W. Lee et al.: Nature \textbf{402}, 55 (1999)

\bibitem{journal} L.R. Bedin et al.: ApJ \textbf{605}, L125 (2004)

\bibitem{journal} J.E. Norris: ApJ \textbf{612}, L25 (2004)

\bibitem{journal} Y.-W. Lee et al.: ApJ \textbf{621}, L57 (2005)

\bibitem{journal} F. D'Antona et al.: ApJ \textbf{631}, 868 (2005)

\bibitem{journal} G. Piotto et al.: ApJ \textbf{661}, L53 (2007)

\bibitem{journal} K. Bekki \& J.E. Norris: ApJ \textbf{637}, L109 (2006)

\bibitem{journal} Y.-W. Lee, H.B. Gim, \& D.I. Casetti-Dinescu: ApJ \textbf{661}, L49 (2007)

\bibitem{journal} S.-C. Rey et al.: ApJS in press (2007)

\bibitem{journal} S.T. Sohn et al.: AJ \textbf{131}, 866 (2006)


\end{thebibliography}
\end{document}